\documentclass[11pt]{article}

\usepackage[preprint]{acl}
\usepackage{times}
\usepackage{latexsym}
\usepackage[T1]{fontenc}
\usepackage[utf8]{inputenc}
\usepackage{microtype}
\usepackage{inconsolata}

\usepackage{color,xcolor}
\usepackage{epsfig}
\usepackage{graphicx}
\usepackage{lipsum}

\usepackage{array}
\usepackage{booktabs}
\usepackage{colortbl}
\usepackage{multirow}
\usepackage{float}
\usepackage{caption}
\usepackage{adjustbox}
\usepackage{subcaption}
\usepackage{footnote}
\makesavenoteenv{tabular}
\makesavenoteenv{table}
\usepackage{makecell}
\usepackage{tablefootnote}
\usepackage{threeparttable}

\usepackage{amsmath,amsfonts,amssymb,bm}
\usepackage[super]{nth}

\usepackage{changepage}
\usepackage{extramarks}
\usepackage{lastpage}
\usepackage{setspace}
\usepackage{soul}
\usepackage{xspace}
\usepackage{wrapfig}

\usepackage{url}
\usepackage{hyperref}
\hypersetup{colorlinks=True, urlcolor=black}

\usepackage[ruled,vlined]{algorithm2e}
\usepackage{enumitem}
\usepackage{verbatim}
\usepackage{pifont}
\usepackage[acronyms]{glossaries}
\glsdisablehyper

\title{UAT: Unified Audio-Text Diffusion for Audio Generation, Editing, and Captioning}

\author{
\textbf{Hui Wang\textsuperscript{1, 2}}\thanks{Work done during internship at Tencent.},
\textbf{Yifan Yang\textsuperscript{3}},
\textbf{Zeyue Tian\textsuperscript{4, 5}},
\textbf{Yuhang Jia\textsuperscript{1}},
\textbf{Jinghua Zhao\textsuperscript{1}},
\textbf{Long Zhou\textsuperscript{2}}\thanks{Project Leader\quad$^{\ddagger}$Corresponding Author}\\
\textbf{Bing Han\textsuperscript{3}},
\textbf{Cheng Liu\textsuperscript{1}},
\textbf{Jiaming Zhou\textsuperscript{1}},
\textbf{Geng Tu\textsuperscript{2}},
\textbf{Yong Qin\textsuperscript{1$\ddagger$}}\\
\textsuperscript{1}College of Computer Science, Nankai University
\textsuperscript{2}Tencent\\
\textsuperscript{3}Shanghai Jiao Tong University
\textsuperscript{4}HKUST
\textsuperscript{5}Noiz AI\\
\small{
\textbf{Correspondence:} \href{mailto:wanghui_hlt@mail.nankai.edu.cn}{wanghui\_hlt@mail.nankai.edu.cn}, \href{mailto:qinyong@nankai.edu.cn}{qinyong@nankai.edu.cn}}
}

\begin{document}
\maketitle

\begin{abstract}
Audio generation and audio-to-text understanding remain largely separate, with diffusion models dominating high-fidelity synthesis and autoregressive (AR) language models driving captioning and semantic prediction. Existing unified approaches typically rely on either heterogeneous modules or AR-centric modeling, which can hinder joint optimization and limit acoustic fidelity. We present UAT, to our knowledge, the first diffusion-centric framework that supports unified audio generation, editing, and captioning. UAT couples continuous latent diffusion for audio with masked discrete diffusion for text, enabling bidirectional audio-text modeling within a shared dual-stream backbone. Experiments show that UAT preserves strong audio generation and editing capabilities while achieving competitive captioning performance, demonstrating a favorable balance between acoustic synthesis and semantic prediction. Demo samples are available at \url{https://UAT-demo.github.io}.

\end{abstract}

\section{Introduction}

Unifying generation and understanding within a single model has emerged as an important research direction. Recent studies have made substantial progress within the image and video domains~\citep{transfusion, zhao2025unified, dualdiffusion, showo}. These works suggest that a unified formulation can facilitate cross-task knowledge transfer, reduce redundant task-specific designs, and bridge the gap between perceptual synthesis and semantic understanding.

Despite these advances, audio generation and audio understanding are still largely studied under separate modeling paradigms. High-fidelity text-to-audio (TTA) generation and editing are predominantly driven by diffusion-based models operating in continuous latent spaces~\citep{audioldm, audioldm2, lafma}, whereas audio captioning and understanding are typically formulated as autoregressive (AR) generation tasks within large language models~\citep{midashenglm, af3}. This separation prevents different tasks from sharing model architectures, representations, and supervision. Consequently, audio synthesis and textual prediction remain optimized in isolation, limiting cross-task transfer, data-efficient learning, and unified audio-text modeling.

\begin{figure}[t]
  \includegraphics[width=\linewidth]{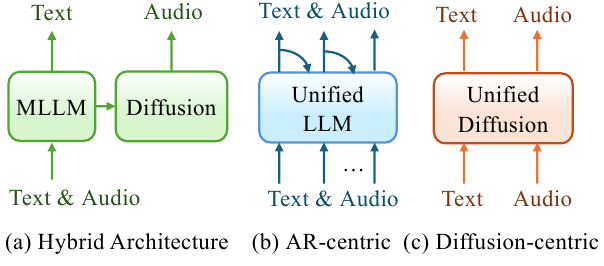}
  \vspace{-15pt}
  \caption{Three architectural routes toward unified audio-text modeling: hybrid systems, autoregressive-centric models, and our diffusion-centric model. }
  \label{fig:structure_overview}
\end{figure}

Recent efforts toward unified audio generation and understanding generally follow two paradigms, as shown in Figure~\ref{fig:structure_overview}. Hybrid architectures~\citep{audioomni} connect frozen multimodal large language models (MLLMs) with diffusion backbones via feature projection, but their generation and understanding components still operate in separate latent spaces and are optimized with different objectives, limiting the ability to jointly model semantic reasoning and acoustic synthesis. AR-centric models~\citep{ualm, uniaudio2, UnifiedIO2} provide a unified sequence-modeling interface by interleaving text tokens with audio representations and predicting discrete audio tokens for synthesis. However, generation quality is limited by the information bottleneck of discrete acoustic tokenization. In addition, strictly left-to-right decoding makes it difficult to correct earlier errors and maintain global acoustic consistency.

These limitations motivate a diffusion-centric alternative for unified audio-text modeling. However, adapting existing text-to-audio diffusion backbones to this setting is non-trivial. \textbf{At the architectural level, current TTA diffusion models are inherently asymmetric}: audio latents are iteratively updated by the diffusion transformer, while text remains a static condition injected through cross-attention. This design lacks an active text stream that can be progressively refined for audio-to-text generation. \textbf{At the modeling level, audio and text exhibit a paradigm discrepancy}: audio synthesis is performed in continuous latent spaces, whereas text generation requires discrete token prediction. Together, these challenges make it difficult to directly repurpose existing TTA diffusion backbones for unified audio generation and captioning.

To address these challenges, we present \textbf{U}nified \textbf{A}udio-\textbf{T}ext Diffusion (\textbf{UAT}), a diffusion framework for audio generation, editing, and captioning. To resolve the architectural asymmetry, UAT extends a pretrained text-to-audio diffusion backbone with a lightweight text stream, forming a coupled dual-stream architecture. To bridge the paradigm discrepancy, UAT combines continuous latent diffusion for acoustic modeling with masked discrete diffusion for textual token generation. Experiments show that this retrofitted unified model achieves strong performance in audio generation and editing while maintaining competitive audio captioning results, supporting the viability of diffusion-centric unified audio-text modeling. Our contributions are summarized as follows:
\begin{itemize}[leftmargin=*, itemsep=0pt, topsep=2pt]
    \item We formulate audio generation, editing, and captioning within a unified diffusion-centric framework, providing a non-autoregressive alternative to unified audio-text modeling.
    \item We introduce a coupled dual-stream architecture that combines continuous audio diffusion with discrete text diffusion, addressing both architectural asymmetry and paradigm discrepancy.
    \item We demonstrate through extensive experiments that UAT preserves strong generation and editing capability while achieving competitive captioning performance.
\end{itemize}

\section{Related Work}

\subsection{Audio Generation and Editing}

Audio generation has achieved substantial progress with the development of diffusion-based models. Recent text-to-audio systems typically perform denoising in continuous waveform or latent spaces, enabling the synthesis of high-fidelity audio that is semantically aligned with textual prompts~\citep{audioldm, audioldm2, audiox, lafma}. Compared with autoregressive generation over discrete audio tokens, diffusion models provide an iterative refinement process that is well-suited for modeling fine-grained acoustic details and complex temporal structures. This property also makes them effective for audio editing~\citep{audit,audioeditor}, where the model is required to modify specific acoustic content while preserving the surrounding context.

Despite their strong synthesis and editing capabilities, existing diffusion-based audio models are mostly designed as one-way conditional generators~\citep{audioldm,tango,stableaudioopen}. Text is usually encoded as a condition and injected into the denoising network through cross-attention or similar mechanisms, while only audio latents are progressively updated. Such an asymmetric formulation is effective for text-conditioned generation, but it does not naturally support inverse tasks such as audio captioning.

\subsection{Audio Captioning and Understanding}

Audio captioning and understanding focus on extracting semantic information from acoustic signals and expressing it in natural language. Recent methods commonly formulate these tasks as audio-to-text generation problems, where an audio encoder first maps input audio into continuous features or discrete representations, and an autoregressive language model then generates captions, answers, or instructions~\citep{af3, qwen2audio,zhao2025omniclst}. Benefiting from the reasoning and language generation ability of large language models, these approaches have shown strong performance on audio captioning, audio question answering, and instruction-following tasks.

However, audio understanding models are optimized for semantic prediction rather than acoustic synthesis. As a result, they are not naturally equipped with the ability to generate or edit high-fidelity audio. To extend such models to generative tasks, prior methods often rely on discrete audio token prediction or external generation modules~\citep{audiogpt}, which can introduce quantization loss, complicate the overall system, and degrade acoustic realism~\citep{guo2025recent,felle}.

\subsection{Unified Audio-Text Modeling}

Recent efforts have explored unified audio-text modeling to bridge audio understanding and generation. One line of work adopts hybrid architectures that connect language models with dedicated audio encoders and audio generators through feature projection or intermediate representations~\citep{audioomni}. By leveraging specialized modules for different modalities and tasks, these systems can support a broad range of audio-text interactions, including speech understanding, audio captioning, and text-guided generation. However, the understanding and generation components often operate in separate representation spaces and are optimized with different objectives, which limits end-to-end cross-modal alignment, joint optimization, and knowledge sharing across tasks.

Another line of work formulates audio-text modeling as autoregressive sequence prediction over text tokens and discrete audio tokens, particularly on the generation side~\citep{ualm,uniaudio2,UnifiedIO2}. By converting audio signals or spectrograms into token sequences, these methods provide a unified interface for audio generation, continuation, captioning, and multimodal reasoning. Despite this conceptual simplicity, discrete audio tokenization can introduce a quantization bottleneck and may discard fine-grained acoustic details that are important for perceptual quality. In addition, autoregressive decoding produces audio tokens sequentially, which can be inefficient for long audio sequences and does not naturally provide the iterative refinement mechanism that is central to diffusion-based synthesis and editing.

\begin{figure}[t]
  \includegraphics[width=\columnwidth]{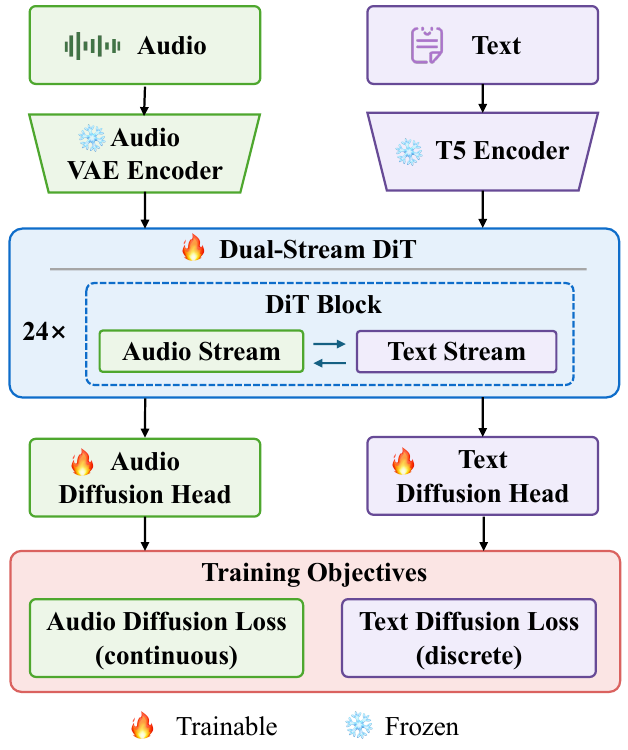}
  \caption{
Overview of UAT, which couples continuous audio diffusion with masked text diffusion in a dual-stream DiT.
}
  \label{fig:architecture}
\end{figure}

\begin{figure*}[t]
\begin{center}
\includegraphics[width=0.98\linewidth]{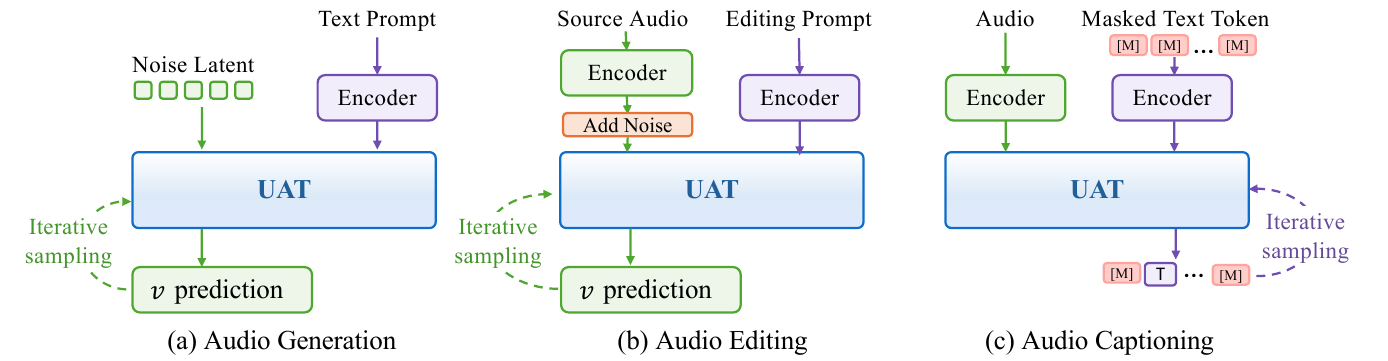}
\end{center}

\caption{
Multi-task inference with UAT. The same dual-stream DiT model supports audio generation, instruction-guided audio editing, and audio captioning by changing the observed condition and the corrupted modality.
}

\label{fig:inference}
\end{figure*}

\section{Method}

\subsection{Problem Formulation}

Given an audio-text pair $(a, y)$, where $a$ is an audio and $y$ is a text sequence, UAT supports three tasks: text-to-audio generation, text-guided audio editing, and audio captioning. We view both audio synthesis and text generation as conditional denoising processes over different modalities. Audio is modeled through continuous latent diffusion, while text is modeled through masked discrete diffusion. Under this view, different tasks correspond to different choices of observed conditions and corrupted target variables, enabling generation, editing, and captioning within a unified diffusion framework.

\subsection{Model Architecture}

As shown in Figure~\ref{fig:architecture}, the model consists of frozen modality encoders, a trainable dual-stream DiT, and two modality-specific output heads.


\paragraph{Modality encoders.}
UAT uses modality-specific encoders to obtain audio and text representations. Given an audio waveform $a$, a frozen audio VAE $E_a$ maps it into a continuous latent representation:
\[
z_0 = E_a(a).
\]
A frozen T5 encoder $E_t$ maps the task-specific text input $y$ into token-level representations:
\[
h^{(0)} = E_t(y).
\]
Here, $y$ is the clean prompt for audio generation, and the corrupted caption for audio captioning.

\paragraph{Dual-stream DiT.}

The core of UAT is a dynamic dual-stream Diffusion Transformer, which maintains an audio stream and a text stream throughout the backbone. The audio stream processes continuous audio latent representations, while the text stream processes token-level text representations. Unlike conventional text-to-audio diffusion models that use fixed text embeddings as conditions, UAT updates both audio and text states layer by layer.

Let $z^{(l)}$ and $h^{(l)}$ denote the audio and text states at the $l$-th layer, respectively. Let $F_a^{(l)}$ and $F_t^{(l)}$ denote the corresponding audio-stream and text-stream update functions in the $l$-th dual-stream DiT layer. The layer-wise interaction is formulated as:
\[
z^{(l+1)} = F_{a}^{(l)}\left(z^{(l)}, h^{(l)}\right),
\]
\[
h^{(l+1)} = F_{t}^{(l)}\left(h^{(l)}, z^{(l+1)}\right).
\]
Here, the audio stream is conditioned on the current text states, and the text stream is conditioned on the updated audio states. Through this mutual conditioning, audio and text representations are dynamically refined and co-evolve within the same diffusion backbone.

\paragraph{Diffusion Heads.}

UAT uses two diffusion heads on top of the dual-stream DiT. The audio diffusion head is inherited from the pretrained backbone and predicts the continuous velocity target for audio denoising, supporting both generation and editing. The text diffusion head maps the final text states to vocabulary logits for masked token reconstruction, with lightweight refiner blocks used to further refine text-side representations before prediction. Together, these two heads enable the same backbone to support continuous audio diffusion and discrete text diffusion.

\subsection{Training Objectives}

UAT is optimized with a joint objective that combines continuous audio diffusion and masked discrete text diffusion. 

\paragraph{Audio diffusion objective.}
Following the Stable Audio-style cosine velocity-prediction objective, we train the audio generation branch to denoise continuous audio latents. Given an audio waveform $a$, we encode it into a clean latent representation $z_0=E_a(a)$. We sample Gaussian noise $\epsilon\sim\mathcal{N}(0,I)$ and a timestep $t\in[0,1]$, and construct the noisy latent as
\[
z_t=\alpha_t z_0+\sigma_t\epsilon,
\]
where $\alpha_t=\cos(\pi t/2)$ and $\sigma_t=\sin(\pi t/2)$. The model predicts the corresponding velocity target $v_{\mathrm{target}}=\alpha_t\epsilon-\sigma_t z_0$:
\[
\mathcal{L}_{\mathrm{audio}}
=
\mathbb{E}_{z_0,\epsilon,t}
\left[
\left\|
v_\theta(z_t,y,t)-v_{\mathrm{target}}
\right\|_2^2
\right].
\]

\paragraph{Masked text diffusion objective.}
For audio captioning, we formulate text generation as masked discrete diffusion. Given a caption $y=\{y_i\}_{i=1}^{L}$, we sample a text diffusion timestep $\tau \in (0,1]$ and independently mask each token with probability $p_{\mathrm{mask}}(\tau)=(1-\varepsilon)\tau$, producing a corrupted caption $y_\tau$. Let $m_i\in\{0,1\}$ indicate whether the $i$-th token is masked. The corrupted caption is processed by the text stream together with the audio latent $z_0$, and the model is trained to reconstruct the original tokens at the masked positions:
\[
\mathcal{L}_{\mathrm{text}}
=
\mathbb{E}_{z_0,y,\tau,m}
\left[
\frac{w(\tau)}{L}
\sum_{i=1}^{L}
m_i \ell_i
\right],
\]
where $\ell_i=-\log p_\theta(y_i\mid c_\tau)$, $c_\tau=(y_\tau,z_0)$, and $w(\tau)=\sigma'(\tau)/(\exp(\sigma(\tau))-1)$ with $\sigma(\tau)=-\log(1-(1-\varepsilon)\tau)$. The clean caption is used only to construct the corrupted input and provide reconstruction targets.

\paragraph{Joint objective.}
The final training objective is:
\[
\mathcal{L}
=
\mathcal{L}_{\mathrm{audio}}
+
\lambda \mathcal{L}_{\mathrm{text}},
\]
where $\lambda$ is a balancing hyperparameter that coordinates generative audio synthesis and text reconstruction. 
This joint optimization enables the dual-stream DiT to learn shared, bidirectional audio-text representations.

\subsection{Multi-Task Inference}
During inference, a single set of trained UAT weights can be flexibly deployed across three major audio-language tasks by activating the corresponding processing pathways, as illustrated in Figure~\ref{fig:inference}.

\paragraph{Audio Generation.}
To perform text-to-audio generation, given an input text prompt $y$, we extract its conditioning representation using the text encoder. Starting from randomly sampled Gaussian audio latents, the audio stream follows the learned velocity field conditioned on the text representation and progressively transports the latent trajectory toward the clean audio latent $\hat{z}_0$. Finally, the frozen VAE decoder reconstructs $\hat{z}_0$ into the output waveform.

\paragraph{Audio Editing.}
For text-guided audio editing, we leverage an SDEdit-style procedure in the continuous latent space~\citep{sdedit}. Given a source audio $a_{\mathrm{src}}$ and a target editing prompt $y_{\mathrm{new}}$, the source audio is first mapped to the latent space as $z_0 = \mathcal{E}_a(a_{\mathrm{src}})$.
We then perturb $z_0$ to an intermediate noise level by adding Gaussian noise:
\[
    z_{t_0} = z_0 + \sigma_{t_0}\epsilon,
    \quad
    \epsilon \sim \mathcal{N}(0,\mathbf{I}),
\]
where $\sigma_{t_0}$ is determined by the inference scheduler, and $t_0$ controls the trade-off between preserving the source audio structure and following the target prompt. Starting from the perturbed latent $z_{t_0}$, UAT follows the learned velocity field under the target text condition $y_{\mathrm{new}}$ to obtain the edited latent $\hat{z}'_0$. The edited waveform is then reconstructed by the frozen VAE decoder.

\paragraph{Audio Captioning.}
For audio-to-text generation, the input audio is first encoded into a continuous latent representation $z_0$ by the frozen audio VAE encoder. The audio latent is processed by the audio stream and provides audio-conditioned features to the text stream. The text sequence is initialized with fully masked tokens. UAT then performs discrete reverse diffusion over text tokens. At each step, the caption head predicts token distributions conditioned on the current partially reconstructed text sequence and the audio latent. The process progressively reconstructs the masked tokens and outputs the final natural-language description.

\section{Experiments}

\subsection{Dataset Description}

We construct a large-scale audio training corpus by integrating multiple public audio sources, including AudioSetCaps~\citep{audiosetcaps}, VGGSound~\citep{vggsound}, AudioCaps~2.0\footnote{\url{https://github.com/cdjkim/audiocaps/tree/master/dataset2.0}}, and WavCaps~\citep{wavcaps}. The resulting corpus contains approximately 2.4M audio samples, totaling about 6.6K hours of audio. Detailed statistics are provided in Appendix~\ref{app:traindata}.

\subsection{Implementation Details}

\paragraph{Model Architecture.}
UAT is initialized from the pretrained AudioX checkpoint\footnote{\url{https://huggingface.co/HKUSTAudio/AudioX}}, which follows the Stable Audio DiT architecture. The DiT backbone contains 24 transformer blocks with a hidden dimension of 1536. The frozen VAE compresses audio into continuous latent representations, while the frozen T5-Base encoder provides 768-dimensional text features. Text branch refines text features in selected DiT blocks via audio-conditioned cross-attention and a residual feed-forward layer.

\paragraph{Training Details.}
For classifier-free guidance (CFG), text conditioning is dropped with a probability of 0.1 during training. The loss balancing weight is set to $\lambda=0.2$. We train the model for 60,000 steps on 32 NVIDIA H20 GPUs using AdamW with a learning rate of $8\times10^{-5}$ and a global batch size of 768. 


\paragraph{Inference Details.}
For audio generation, we use 100 flow-matching sampling steps with a CFG scale of 7.0. For audio editing, we use the same 100-step flow-matching sampler with a CFG scale of 7.0, and start the editing trajectory from step 70. 


\begin{table*}[t]
\centering
\small
\setlength{\tabcolsep}{3.5pt}
\resizebox{\textwidth}{!}{
\begin{tabular}{llccccc|ccccc}
\toprule
\multirow{2}{*}{\textbf{Model Type}} 
& \multirow{2}{*}{\textbf{Model}} 
& \multicolumn{5}{c|}{\textbf{AudioCaps test set}} 
& \multicolumn{5}{c}{\textbf{VGGSound test set}} \\
\cmidrule(lr){3-7} \cmidrule(lr){8-12}
& 
& \textbf{KL $\downarrow$} 
& \textbf{IS $\uparrow$} 
& \textbf{FD $\downarrow$} 
& \textbf{FAD $\downarrow$} 
& \textbf{CLAP $\uparrow$}
& \textbf{KL $\downarrow$} 
& \textbf{IS $\uparrow$} 
& \textbf{FD $\downarrow$} 
& \textbf{FAD $\downarrow$} 
& \textbf{CLAP $\uparrow$} \\
\midrule

\multirow{6}{*}{\makecell{Specialized\\Models}}
& Tango 2          & 1.12 & 10.65 & 11.55 & 2.82  & 0.568 & 1.48 & 6.21 & 31.01 & 4.33 & 0.337   \\
& AudioLDM        & 1.98 & 6.67  & 34.71 & 8.01  & 0.355 & 1.49  & 6.41   & 35.66   & 9.88  & 0.432  \\
& AudioLDM 2       & 1.46 & 9.45  & 17.66 & 1.83  & 0.444 &1.17  &6.96  &19.65 &6.32  &0.380    \\
& MAGNeT          & 1.69 & 6.90  & 27.09 & 3.12  & 0.380 & 1.28 & 6.12 & 28.80 & 4.80  & 0.335 \\
& Stable Audio Open & 2.74 & 7.37  & 41.45 & 8.83  & 0.211 & 1.89 & 6.67 & 39.25 & 7.75  & 0.304 \\
& AudioX          & 1.37 & 12.05 & 13.03 & 2.03  & 0.488 & 1.29 & 8.97 & 21.09 & 5.31  & 0.439 \\
\midrule

\multirow{4}{*}{Unified Models}

& Unified-IO 2     & 2.79 & 4.12  & 82.54 & 21.88 & 0.189 & 2.25 & 3.96 & 80.94 & 21.02 & 0.174 \\
& UniAudio 2.0     & 3.25 & 4.81  & 53.55 & 9.99  & 0.087 & 2.69 & 5.34 & 49.39 & 10.25 & 0.151 \\
& Audio-Omni       & \textbf{1.39} & 9.94  & 45.43 & \textbf{2.00}  & \textbf{0.498} & 1.33 & 8.31 & 53.97 & \textbf{4.56}  & 0.407 \\
& Ours            & \textbf{1.39} & \textbf{12.47} & \textbf{14.47} & 2.87  & 0.491    & \textbf{1.28}   & \textbf{9.34}   & \textbf{22.07}    & 4.91    & \textbf{0.434}  \\

\bottomrule
\end{tabular}
}
\caption{Comparison of text-to-audio generation performance on the AudioCaps and VGGSound test sets. Bold numbers indicate the best results among unified models.}
\label{tab:generation}
\vspace{-5pt}
\end{table*}

\begin{table}[t]
\centering
\small
\setlength{\tabcolsep}{8pt}
\begin{tabular}{lcc}
\toprule
\textbf{Model} 
& \textbf{OVL $\uparrow$} 
& \textbf{REL $\uparrow$} \\
\midrule

Ground Truth & $4.347 \pm 0.142$ & $4.407 \pm 0.157$ \\
\midrule

Unified-IO 2  & $2.853 \pm 0.200$ & $2.967 \pm 0.214$ \\
UniAudio 2.0  & $3.620 \pm 0.171$ & $3.160 \pm 0.180$ \\
Audio-Omni   & $4.047 \pm 0.133$ & $3.893 \pm 0.157$ \\
Ours         & $\mathbf{4.260 \pm 0.131}$ & $\mathbf{4.260 \pm 0.155}$ \\

\bottomrule
\end{tabular}
\caption{Human evaluation results on overall quality (OVL) and relevance (REL). }
\label{tab:human_eval}
\vspace{-5pt}
\end{table}

\begin{table*}[t]
\centering
\small
\setlength{\tabcolsep}{3.5pt}
\resizebox{\textwidth}{!}{
\begin{tabular}{llccc|ccc|ccc}
\toprule
\multirow{2}{*}{\textbf{Model Type}} 
& \multirow{2}{*}{\textbf{Model}} 
& \multicolumn{3}{c|}{\textbf{Add}} 
& \multicolumn{3}{c|}{\textbf{Delete}} 
& \multicolumn{3}{c}{\textbf{Replace}} \\
\cmidrule(lr){3-5} \cmidrule(lr){6-8} \cmidrule(lr){9-11}
& 
& \textbf{CLAP $\uparrow$} 
& \textbf{FAD $\downarrow$} 
& \textbf{IS $\uparrow$}
& \textbf{CLAP $\uparrow$} 
& \textbf{FAD $\downarrow$} 
& \textbf{IS $\uparrow$}
& \textbf{CLAP $\uparrow$} 
& \textbf{FAD $\downarrow$} 
& \textbf{IS $\uparrow$} \\
\midrule

\multirow{4}{*}{Specialized Models}
& AP‑adapter      & 0.387 & 45.683 & 4.138 & 0.401 & 48.148 & 3.088 & 0.432 & 47.309 & 4.117 \\
& CycleDiffusion & 0.434 & 4.671  & 3.451 & 0.355 & 3.516  & 2.867 & 0.447 & 5.968  & 3.071 \\
& DDIM Inversion & 0.384 & 4.348  & 3.266 & 0.316 & 5.736  & 2.544 & 0.385 & 6.111  & 2.844 \\
& MusicGen       & 0.382 & 2.599  & 3.646 & 0.342 & 4.284  & 3.126 & 0.404 & 4.230  & 3.731 \\
\midrule

\multirow{2}{*}{Unified Models}
& Audio-Omni     & 0.326 & 45.378 & 3.422 & 0.255 & 48.172 & 2.167 & 0.317 & 47.195 & \textbf{4.147} \\
& Ours           & \textbf{0.406} & \textbf{3.220}  & \textbf{4.072} & \textbf{0.350} & \textbf{4.243}  & \textbf{3.325} & \textbf{0.439} & \textbf{5.199}  & 3.682 \\
\bottomrule
\end{tabular}
}
\caption{Editing performance comparison under Add, Delete, and Replace settings. Bold numbers indicate the best performance among unified models.}

\label{tab:editing}
\vspace{-5pt}
\end{table*}
\begin{table*}[t]
\centering
\small
\setlength{\tabcolsep}{4pt}
\resizebox{\textwidth}{!}{
\begin{tabular}{llcccccc}
\toprule
\textbf{Model Type} & \textbf{Method} & \textbf{Params} 
& \textbf{CIDEr} 
& \textbf{SPICE} 
& \textbf{SPIDEr} 
& \textbf{SBERT-SIM} 
& \textbf{FENSE} \\
\midrule

\multirow{5}{*}{Specialized Models}
& MiDashengLM    & 7.6B                     & 0.397 & 0.133 & 0.265 & 0.583 & 58.04 \\
& Qwen2-Audio       & 8.2B                   & 0.206 & 0.080 & 0.143 & 0.412 & 36.82 \\
& Qwen3-Omni        & 34.5B        & 0.270 & 0.131 & 0.200 & 0.559 & 54.66 \\
& Audio Flamingo 2  & 4.7B             & 0.418 & 0.112 & 0.265 & 0.503 & 49.30 \\
& Audio Flamingo 3  & 9B            & 0.614 & 0.184 & 0.399 & 0.635 & 63.36 \\
\midrule

\multirow{4}{*}{Unified Models}
& Unified-IO 2       & 1.1B                    & 0.112 & 0.069 & 0.090 & 0.379 & 37.64 \\
& UniAudio 2.0       & 4.9B                  & \textbf{0.603} & \textbf{0.147} & \textbf{0.375} & \underline{0.571} & \textbf{56.06} \\
& Audio-Omni         & 7.9B & 0.167 & 0.131 & 0.149 & 0.555 & 48.89 \\
& Ours              & 1.7B                    & \underline{0.406} & \underline{0.139} & \underline{0.272} & \textbf{0.572} & \underline{54.08} \\

\bottomrule
\end{tabular}
}
\vspace{-5pt}
\caption{Performance comparison on audio captioning metrics. Bold numbers indicate the best performance among unified models, and underlined numbers indicate the second-best performance.}
\label{tab:captioning}
\vspace{-5pt}
\end{table*}

\begin{table*}[t]
\centering
\small
\setlength{\tabcolsep}{3.5pt}
\resizebox{\textwidth}{!}{
\begin{tabular}{lccccc|ccccc}
\toprule
\multirow{2}{*}{\textbf{Pretrain Model}} 
& \multicolumn{5}{c|}{\textbf{Audio Generation}} 
& \multicolumn{5}{c}{\textbf{Audio Caption}} \\
\cmidrule(lr){2-6} \cmidrule(lr){7-11}
& \textbf{KL $\downarrow$}
& \textbf{IS $\uparrow$}
& \textbf{FD $\downarrow$}
& \textbf{FAD $\downarrow$}
& \textbf{CLAP $\uparrow$}
& \textbf{CIDEr $\uparrow$}
& \textbf{SPICE $\uparrow$}
& \textbf{SPIDEr $\uparrow$}
& \textbf{SBERT-SIM $\uparrow$}
& \textbf{FENSE $\uparrow$} \\
\midrule

AudioX   & \textbf{1.39} & \textbf{12.47} & \textbf{14.47} & \textbf{2.87} & \textbf{0.49} & \textbf{0.41} & \textbf{0.14} & \textbf{0.27} & \textbf{0.57} & \textbf{54.08} \\
Stable Audio Open & 2.33          & 8.10           & 27.51          & 7.08  & 0.29  & 0.38 & 0.12 & 0.25 & 0.55          & 52.26  \\

\bottomrule
\end{tabular}
}
\vspace{-5pt}
\caption{Effect of different pre-trained audio diffusion backbones on audio generation and captioning performance.}
\label{tab:ablation_pretrain}
\vspace{-5pt}
\end{table*}

\subsection{Evaluation}

\paragraph{Evaluation Datasets.}
For audio generation, we evaluate on the AudioCaps~\citep{audiocaps} and VGGSound test sets. For audio editing, we evaluate on the Add, Delete, and Replace settings from AuditScore-Bench~\citep{editeval}. For audio captioning, we report results on AudioCaps captioning benchmarks. A detailed description of the editing data is provided in Appendix~\ref{app:testset}.

\paragraph{Evaluation Metrics.}
For audio generation, we evaluate model performance using both objective and subjective evaluations. For objective evaluation, we report KL divergence, Inception Score (IS)~\citep{is}, Fréchet Distance (FD)~\citep{fd}, Fréchet Audio Distance (FAD)~\citep{fad}, and CLAP score~\citep{clap}~\footnote{\url{https://github.com/LAION-AI/CLAP}}. KL divergence, FD, and FAD quantify the distributional discrepancy between generated and reference audio. IS measures the diversity of generated audio, while CLAP evaluates text-audio semantic alignment. For subjective evaluation, we randomly sample 30 examples from the test set and conduct human evaluation along two dimensions: overall quality and relevance to the text prompt. Each audio sample is rated by five human evaluators on a 1--5 scale. Details are provided in Appendix~\ref{app:mostest}.

For audio editing, we use the CLAP score between the edited audio and the target prompt as a proxy for instruction-following. We further report FAD and IS relative to the source audio to assess the preservation of distributional and perceptual characteristics from the original audio.

For audio captioning, we report CIDEr~\citep{cider}, SPICE~\citep{spice}, SPIDEr~\citep{spider}, SBERT similarity (SBERT-SIM)~\citep{sbertsim}, and FENSE~\citep{fense}, computed using the \texttt{aac-metrics} package\footnote{\url{https://github.com/Labbeti/aac-metrics}}, to assess both lexical overlap and semantic-level caption quality.

\paragraph{Baselines.}

We compare UAT with specialized task-specific baselines and unified audio-text models. Specialized baselines include Tango~2~\citep{tango2}, AudioLDM~\citep{audioldm}, AudioLDM~2~\citep{audioldm2}, MAGNeT~\citep{magnet}, Stable Audio Open~\citep{stableaudioopen}, and AudioX~\citep{audiox} for audio generation; AP-adapter~\citep{adapter}, CycleDiffusion~\citep{cyclediffusion}, DDIM Inversion~\citep{ho2020denoising, song2020denoising}, and MusicGen~\citep{musicgen} for audio editing; and MiDashengLM~\citep{midashenglm}, Qwen2-Audio~\citep{qwen2audio}, Qwen3-Omni~\citep{qwen3omni}, Audio Flamingo~2~\cite{af2}, and Audio Flamingo~3~\cite{af3} for audio captioning. Unified baselines include Unified-IO 2~\citep{UnifiedIO2}, UniAudio 2.0~\citep{uniaudio2}, and Audio-Omni~\citep{audioomni}, with only Audio-Omni evaluated on audio editing because the other unified models do not support this task. See Appendix~\ref{app:baseline} for details.

\section{Results}

\subsection{Main Results}

\paragraph{Audio Generation.}
Table~\ref{tab:generation} reports text-to-audio generation results on the AudioCaps and VGGSound test sets. Compared with existing unified audio-text models, UAT achieves the strongest overall generation performance. On AudioCaps, UAT obtains the best IS among all evaluated models and substantially outperforms Unified-IO 2 and UniAudio 2.0 across all metrics. Compared with Audio-Omni, UAT achieves much higher IS and significantly lower FD, while maintaining comparable KL and CLAP scores. On VGGSound, UAT also achieves the best IS and the lowest KL among unified models, with FD and CLAP scores close to the specialized AudioX model. These results indicate that introducing a text stream for captioning does not destroy the generation capability of the pretrained diffusion backbone. Although some specialized TTA models remain strong on specific metrics, UAT achieves a favorable balance between generation quality and unified modeling ability, showing that a diffusion-centric unified model can retain competitive audio synthesis performance.

Table~\ref{tab:human_eval} further presents human evaluation results on overall quality and relevance. UAT achieves an OVL score of 4.260 and a REL score of 4.260, which are close to the ground-truth scores of 4.347 and 4.407. UAT also clearly outperforms other unified models, including Unified-IO 2, UniAudio 2.0, and Audio-Omni. This confirms that the generated audio is not only strong under automatic metrics, but also preferred by human listeners in terms of perceptual quality and text relevance.

\paragraph{Audio Editing.}
Table~\ref{tab:editing} compares audio editing performance under the Add, Delete, and Replace settings. Compared with the unified baseline Audio-Omni, UAT achieves consistently higher CLAP scores and lower FAD across all three editing scenarios, indicating better instruction following and stronger preservation of the original audio distribution. UAT also improves IS under the Add and Delete settings, although Audio-Omni obtains a higher IS under Replace.

Compared with specialized editing baselines, UAT does not always achieve the best score on individual metrics, but it shows a more balanced performance across semantic alignment, audio quality, and distributional fidelity.
In particular, UAT avoids the severe FAD degradation observed in some methods, while maintaining competitive CLAP and IS scores across Add, Delete, and Replace. These results suggest that UAT can support flexible audio editing within a unified diffusion framework, achieving a favorable trade-off between controllability, quality, and generality.

\paragraph{Audio Captioning.}
Table~\ref{tab:captioning} reports audio captioning results. Compared with unified audio-text baselines, UAT achieves competitive captioning performance with a moderate model size. UAT substantially outperforms Unified-IO 2 and Audio-Omni on CIDEr, SPIDEr, SBERT similarity, and FENSE, and obtains the highest SBERT similarity among unified models. Although UniAudio 2.0 achieves higher CIDEr, SPICE, and SPIDEr, UAT uses fewer parameters and maintains much stronger audio generation and editing performance, making it a more balanced unified audio-text model.

Compared with specialized understanding models, UAT is competitive with several large audio-language models despite being optimized under a unified diffusion framework. It outperforms Qwen2-Audio and Qwen3-Omni on multiple captioning metrics, and achieves results comparable to MiDashengLM and Audio Flamingo 2 on CIDEr and SPIDEr. Nevertheless, UAT still lags behind Audio Flamingo 3, reflecting that its understanding capability remains an area for future improvement. Overall, these results suggest that masked discrete diffusion is a viable mechanism for audio-conditioned text generation within a diffusion-centric unified audio-text model.

\begin{figure}[t]
  \includegraphics[width=\columnwidth]{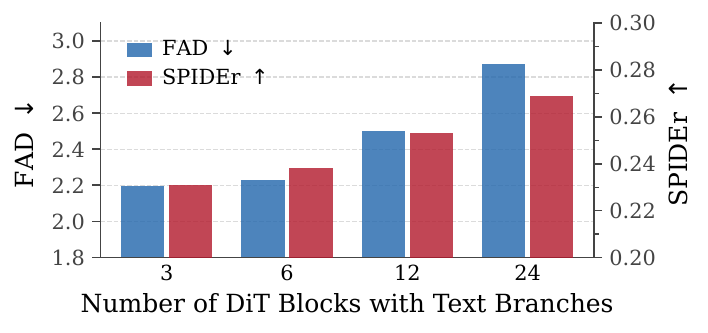}
  \vspace{-20pt}
  \caption{Effect of text-branch depth on audio generation and captioning performance.}
  
  \label{fig:text_branch_depth}
  \vspace{-5pt}
\end{figure}

\subsection{Ablation Studies}

We conduct ablation studies on key design choices of UAT to study their effects on generation and understanding capabilities, with additional results provided in Appendix~\ref{app:res}.

\paragraph{Effect of text branch depth.}
We first investigate how the depth of the inserted text branch affects unified audio-text modeling. Here, depth refers to the number of DiT blocks augmented with an additional text branch, ranging from all 24 blocks to only the last few blocks. As shown in Figure~\ref{fig:text_branch_depth}, reducing the number of text-branch blocks from 24 to 3 gradually improves audio generation quality, as indicated by lower Audio FAD, but consistently degrades captioning performance, reflected by lower Caption SPIDEr. This reveals a trade-off between preserving the original audio denoising capability and improving text-side semantic modeling.

A deeper text branch provides more capacity for masked token recovery and enables richer audio-text interaction, leading to better captioning performance. In contrast, a shallower text branch perturbs the pretrained audio diffusion pathway less, thereby better preserving audio generation quality. These results suggest that the depth of the text branch should be carefully balanced for unified generation and understanding.

\paragraph{Effect of pretrained audio backbone.}
We further examine the effect of the pretrained audio diffusion backbone. As shown in Table~\ref{tab:ablation_pretrain}, UAT initialized from AudioX consistently outperforms the variant initialized from Stable Audio Open on both audio generation and captioning metrics, indicating that a stronger pretrained text-to-audio diffusion backbone brings larger benefits to unified audio-text modeling. Specifically, the AudioX-based model achieves better KL, IS, FD, FAD, and CLAP scores, suggesting that stronger pre-training provides more effective acoustic priors for audio synthesis. It also improves CIDEr, SPICE, SPIDEr, SBERT similarity, and FENSE, showing that stronger audio representations and generation priors can further benefit audio-conditioned semantic prediction. These results demonstrate that the choice of pretrained diffusion backbone affects not only generation quality but also the effectiveness of text generation in unified audio-text modeling.

\section{Conclusion}

In this paper, we introduced UAT, a diffusion-centric unified audio-text framework built upon a pretrained audio generation backbone. By integrating a dual-stream DiT, UAT jointly supports continuous latent diffusion for audio generation and editing, and masked discrete diffusion for audio captioning. Experiments demonstrate that UAT achieves a favorable balance between acoustic fidelity and semantic understanding. Compared with existing unified models, UAT obtains superior performance on multiple metrics while remaining competitive with task-specific systems. These results highlight the potential of diffusion models not only as powerful audio generators but also as a foundation for unified audio-text modeling.


\section*{Limitations}

Although UAT demonstrates the feasibility of unified audio-text diffusion modeling, it still has several limitations:
(1) UAT relies on the capability of the underlying text-to-audio diffusion backbone. Since our model is built by extending a pretrained audio generation model, its generation and editing quality may still be constrained by the backbone's ability in acoustic realism, prompt following, and coverage of diverse sound events.
(2) The current understanding ability of UAT is still relatively limited compared with large autoregressive audio-language models. While UAT supports audio captioning within the same diffusion-centric framework, tasks requiring complex reasoning, long-form responses, or external knowledge remain challenging.
(3) UAT has not yet been fully explored on broader audio-language tasks. Nevertheless, its unified architecture provides a natural path for future extension. Since new tasks can be introduced by changing the training data rather than modifying the model architecture, UAT can be extended to audio question answering and other audio-language reasoning tasks with corresponding annotated data.

\section*{Ethical Considerations}

This work may benefit audio content creation, automatic audio description, and multimodal accessibility by unifying audio generation, editing, and captioning in a single framework. However, audio generation and editing models may also be misused to create deceptive or unauthorized synthetic audio. Responsible use should include clear disclosure of generated or edited content, respect for consent and copyright, and safeguards such as watermarking or synthetic audio detection. Models may also inherit biases from training data, which calls for careful dataset curation and evaluation.

AI assistants were used in the preparation of this work for data processing and language polishing.


\bibliography{custom}

\clearpage

\appendix

\section{Supplementary Experimental Settings}
\label{sec:appendix}

\subsection{Training Dataset Detail}
\label{app:traindata}

The training corpus consists of four filtered audio-text datasets, as summarized in Table~\ref{tab:training_data}. We filter these datasets to remove any overlap with the evaluation test sets and retain audio clips of approximately 10 seconds. For the VGGSound subset, part of the captions are adopted from the IF-Caps~\citep{audiox} annotations.

We use task-specific sampling ratios for different training objectives. For text-to-audio training, we sample from AudioSetCaps, AudioCaps~2.0, VGGSound, and WavCaps with ratios of 50\%, 20\%, 15\%, and 15\%, respectively. For audio-to-text training, we use AudioSetCaps, AudioCaps~2.0, and WavCaps with ratios of 15\%, 60\%, and 25\%, respectively. This strategy allows generation training to benefit from large-scale and diverse audio-text pairs, while captioning training places more emphasis on higher-quality caption annotations.


\subsection{Audio Editing Test Set Details}
\label{app:testset}
AuditScore-Bench~\citep{editeval} is constructed from the AudioCaps dataset~\citep{audiocaps} and comprises 240 test samples, covering three audio editing operations: Add, Delete, and Replace, with 80 samples for each category. Each sample consists of a real audio clip from AudioCaps paired with triplet annotations, including a natural-language editing instruction (e.g., ``Add a woman talking.''), an original prompt describing the source audio, and a target prompt describing the desired edited audio. Specifically, the Add operation requires the model to introduce a new sound source while preserving the original acoustic events; the Delete operation requires removing a specified sound source without altering the remaining content; and the Replace operation requires substituting a sound source in the original audio with another type of sound source.

\subsection{Subjective Evaluation Protocol}
\label{app:mostest}

We conducted a randomized and anonymous subjective evaluation for audio generation. Audio samples from different systems were pooled together and randomly shuffled before being presented to evaluators. The system identity and file name were hidden, and only the corresponding text prompt was shown. Evaluators were allowed to replay each audio sample as many times as needed before assigning scores on two 1--5 scales: overall quality (OVL) and relevance (REL). OVL measures overall audio quality, including clarity, naturalness, noise, distortion, and perceptual fidelity. REL measures how well the audio matches the text prompt, considering the presence of key sound events, sound sources, background context, temporal order, and relative salience. All ratings were automatically saved and then aggregated by system. We report the mean score across all samples and evaluators, together with 95\% confidence intervals to reflect rating uncertainty.

\begin{table}[t]
\centering
\small
\setlength{\tabcolsep}{8pt}
\resizebox{\columnwidth}{!}{
\begin{tabular}{lcc}
\toprule
\textbf{Dataset} 
& \textbf{\# Samples} 
& \textbf{Duration} \\
\midrule

AudioSetCaps  & 1,993,704 & 5,539.176 h \\
AudioCaps~2.0 & 91,254    & 253.483 h   \\
VGGSound      & 163,759   & 508.383 h   \\
WavCaps       & 115,048   & 319.578 h   \\
\midrule
\textbf{Total} & \textbf{2,363,765} & \textbf{6,620.620 h} \\

\bottomrule
\end{tabular}
}
\caption{Statistics of the training data used in our model.}
\label{tab:training_data}
\end{table}

\begin{table}[t]
\centering
\small
\setlength{\tabcolsep}{5pt}
\begin{tabular}{lccccc}
\toprule
\textbf{Setting} 
& \textbf{KL $\downarrow$} 
& \textbf{IS $\uparrow$} 
& \textbf{FD $\downarrow$} 
& \textbf{FAD $\downarrow$} 
& \textbf{CLAP $\uparrow$} \\
\midrule
Unify        & 1.39 & 12.47 & 14.47 & 2.87 & 0.491 \\
Audio-only & 1.33 & 13.04 & 11.83 & 1.92 & 0.501 \\
\bottomrule
\end{tabular}
\caption{Effect of single-task and unified training on text-to-audio generation. 
The audio-only variant is trained only with the continuous audio diffusion objective, while the unified model is trained jointly. }
\label{tab:single_task_audio}
\end{table}

\begin{table*}[t]
\centering
\setlength{\tabcolsep}{5pt}
\begin{tabular}{lccccc}
\toprule
\textbf{Refiner} 
& \textbf{CIDEr $\uparrow$} 
& \textbf{SPICE $\uparrow$} 
& \textbf{SPIDEr $\uparrow$} 
& \textbf{SBERT-SIM $\uparrow$} 
& \textbf{FENSE $\uparrow$} \\
\midrule
Unify         & 0.406 & 0.139 & 0.272 & 0.572 & 54.08 \\
Caption-only & 0.370 & 0.128 & 0.249 & 0.564 & 53.37 \\
\bottomrule
\end{tabular}
\caption{Effect of single-task and unified training on audio captioning. 
The caption-only variant is trained only with the text diffusion objective, while the unified model is jointly optimized with both audio and text diffusion losses. }
\label{tab:single_task_caption}
\end{table*}
\begin{table*}[t]
\centering
\small
\setlength{\tabcolsep}{3.5pt}
\resizebox{\textwidth}{!}{
\begin{tabular}{lccccc|ccccc}
\toprule
\multirow{2}{*}{\textbf{Refiner}} 
& \multicolumn{5}{c|}{\textbf{Audio Generation}} 
& \multicolumn{5}{c}{\textbf{Audio Captioning}} \\
\cmidrule(lr){2-6} \cmidrule(lr){7-11}
& \textbf{KL $\downarrow$} 
& \textbf{IS $\uparrow$} 
& \textbf{FD $\downarrow$} 
& \textbf{FAD $\downarrow$} 
& \textbf{CLAP $\uparrow$} 
& \textbf{CIDEr $\uparrow$} 
& \textbf{SPICE $\uparrow$} 
& \textbf{SPIDEr $\uparrow$} 
& \textbf{SBERT-SIM $\uparrow$} 
& \textbf{FENSE $\uparrow$} \\
\midrule

1-layer  & 1.41          & 12.75          & 14.29         & 3.00          & 0.482          & 0.380          & 0.131          & 0.255          & 0.558          & 52.98          \\
3-layer  & \textbf{1.39} & 12.47          & \textbf{14.47} & \textbf{2.87} & \textbf{0.491} & \textbf{0.406} & \textbf{0.139} & \textbf{0.272} & 0.572          & 54.08         \\
6-layer  & 1.42         & 12.37          & 14.81          & 3.05          & 0.483          & 0.354          & 0.129          & 0.242          & 0.563          & 50.04          \\
12-layer & 1.41          & \textbf{12.83} & 13.92          & 2.96          & 0.483          & 0.393          & 0.135          & 0.263          & \textbf{0.578} & \textbf{54.92} \\

\bottomrule
\end{tabular}
}
\caption{Ablation results with different numbers of refiners on audio generation and audio captioning tasks.}
\label{tab:refiner_ablation}
\end{table*}

\subsection{Baselines}
\label{app:baseline}

For audio generation, we employ Tango2\footnote{\url{https://huggingface.co/declare-lab/tango2}}, AudioLDM\footnote{\url{https://huggingface.co/cvssp/audioldm}}, AudioLDM 2\footnote{\url{https://huggingface.co/cvssp/audioldm2}}, MAGNeT\footnote{\url{https://huggingface.co/facebook/audio-magnet-medium}}, Stable Audio Open\footnote{\url{https://huggingface.co/stabilityai/stable-audio-open-1.0}}, and AudioX\footnote{\url{https://huggingface.co/HKUSTAudio/AudioX}}. All models are evaluated using their official checkpoints and recommended inference configurations.

For audio editing tasks, we employ four representative generative audio frameworks, namely DDIM Inversion, CycleDiffusion, AP‑adapter\footnote{\url{https://huggingface.co/cvssp/audioldm2}}, and MusicGen\footnote{\url{https://huggingface.co/facebook/musicgen-large}}. We adopt Tango 2\footnote{\url{https://huggingface.co/declare-lab/tango2}} as the foundational generative model for DDIM Inversion and CycleDiffusion, where 125 out of 200 total denoising steps are used for inversion. For AP‑adapter, we adopt AudioLDM 2\footnote{\url{https://huggingface.co/cvssp/audioldm2}} as the foundational generative model.

For audio captioning, we compare our approach against five recent open-source audio-language models. For all baselines, we use the official checkpoints and follow each model's recommended inference configuration. The five baselines are: MiDashengLM\footnote{\url{https://huggingface.co/mispeech/midashenglm-7b-0804-fp32}}, Qwen2-Audio\footnote{\url{https://huggingface.co/Qwen/Qwen2-Audio-7B}}, Qwen3-Omni\footnote{\url{https://huggingface.co/Qwen/Qwen3-Omni-30B-A3B-Instruct}}, Audio Flamingo 2\footnote{\url{https://huggingface.co/nvidia/audio-flamingo-2}}, and Audio Flamingo 3\footnote{\url{https://huggingface.co/nvidia/audio-flamingo-3/}}. MiDashengLM, Qwen2-Audio, and Qwen3-Omni are evaluated using the \texttt{ms-swift}\footnote{\url{https://github.com/modelscope/ms-swift}} framework, while Audio Flamingo 2 and Audio Flamingo 3 are evaluated with their official inference scripts. To ensure a fair comparison, all models are queried with the same user prompt:``\emph{<audio>Write a short caption describing the sounds you hear.''}

\section{Supplementary Experimental Results}
\label{app:res}

\paragraph{Effect of single-task and unified training.}
We further compare UAT with single-task variants trained using only the audio diffusion objective or only the captioning objective. As shown in Table~\ref{tab:single_task_audio}, the audio-only variant achieves better text-to-audio generation metrics than the unified model, indicating that introducing the text diffusion objective slightly perturbs the pretrained audio generation pathway. This is consistent with the trade-off observed in the text-branch depth ablation.

On the other hand, Table~\ref{tab:single_task_caption} shows that the unified model consistently improves over the caption-only variant across all captioning metrics, including CIDEr, SPICE, SPIDEr, SBERT-SIM, and FENSE. These results suggest that joint audio diffusion training can provide useful acoustic-semantic representations for masked text diffusion. Overall, the unified objective does not aim to optimize each task in isolation; instead, it provides a balanced trade-off that enables a single diffusion-centric model to support audio generation, editing, and captioning simultaneously.

\paragraph{Effect of caption refiner.}

We introduce a lightweight caption refiner before the vocabulary projection in the Caption Diffusion Head. The refiner consists of stacked Transformer-style self-attention blocks that refine the resulting text hidden states for caption reconstruction. It improves the expressiveness of the caption head while keeping the pretrained audio backbone reusable. 

As shown in Table~\ref{tab:refiner_ablation}, increasing the refiner depth from 1 to 3 consistently improves both audio generation and captioning metrics, indicating that a moderate number of self-attention refinement layers helps transform the backbone text states into more discriminative representations for the auxiliary captioning objective. However, further increasing the depth does not yield monotonic improvements. The 6-layer refiner performs worse across most metrics, suggesting that an overly deep caption head may introduce optimization difficulty or absorb the caption supervision within the head itself, weakening its regularization effect on the shared audio-text backbone. The 12-layer refiner partially recovers on semantic captioning metrics, but still underperforms the 3-layer refiner on the main audio distribution metrics and CIDEr/SPIDEr. Overall, the 3-layer refiner provides the best balance between sufficient text-side capacity and effective joint audio-caption optimization.

\end{document}